\documentclass[12pt,a4paper]{article}%
\usepackage{amsfonts}
\usepackage{amsmath}
\usepackage{eurosym}
\usepackage{amssymb}
\usepackage{graphicx}%
\providecommand{\U}[1]{\protect\rule{.1in}{.1in}}
\begin{document}

\title{Anti-$\mathcal{PT}$-symmetric harmonic oscillator and its relation to the
inverted harmonic oscillator }
\author{Nadjat Amaouche$^{a\text{,}1}$, Ishak Bouguerche$^{a,2}$, Rahma
Zerimeche$^{a\text{,}b,3}$
\and and Mustapha Maamache$^{a,4}$\thanks{E-mails: $^{\text{1}}$%
nadjatamaouch@gmail.com,$^{\text{2}}$isaacmadrid77@yahoo.com,$^{\text{3}}%
$zerimecherahma@gmail.com,$^{\text{4}}$maamache@univ-setif.dz }\\$^{a}$Laboratoire de Physique Quantique et Syst\`{e}mes Dynamiques,\\Facult\'{e} des Sciences, Ferhat Abbas S\'{e}tif 1, S\'{e}tif 19000, Algeria.\\$^{b\text{ \ }}$Physics Department, University of Jijel, BP 98, Ouled Aissa, \\18000 Jijel, Algeria{\small .}}
\date{}
\maketitle

\begin{abstract}
We treat the quantum dynamics of a harmonic oscillator as well as its inverted
counterpart in the Schr\"{o}dinger picture. Generally in the most papers of
the literature, the inverted harmonic oscillator is formally obtained from the
harmonic oscillator by the replacement$\ $\textsl{\ }of$\ \omega$ to $i\omega
$, this leads to unbounded eigenvectors. This explicitly demonstrates that
there are some unclear points involved in redefining the variables in the
harmonic oscillator inversion. To remedy this situation, we introduce a
scaling operator (Dyson transformation) by connecting the inverted harmonic
oscillator to an anti-$\mathcal{PT}$-symmetric harmonic oscillator, we obtain
the standard quasi-Hermiticity relation which would ensure the time invariance
of the eigenfunction's norm. We give a complete description for the
eigenproblem\textrm{.} We show\textrm{ }that\textrm{ }the wavefunctions for
this system are\textrm{ }normalized in the sense of the pseudo-scalar product.
A Gaussian wave packet of the inverted oscillator is investigated by using the
ladder operators method. This wave packet\textrm{ }is\textrm{ }found to be
associated with the generalized coherent state that can be crucially utilized
for investigating the mean values of the space and momentum operators. We find
that these mean values reproduce the classical motion.


\textbf{This paper is dedicated to the memory of ours colleagues: Brahim
Bouzerafa and Rabah Zegadi who died due to covid 19. And to Ali-Sahraoui
Ferhat due to cardiac arrest.}

\end{abstract}

\section{\bigskip Introduction}

The inverted harmonic oscillator described by the following operator :%
\begin{equation}
H^{\mathrm{r}}=\frac{1}{2m}p^{2}-\frac{1}{2}m\omega^{2}x^{2}, \label{r1}%
\end{equation}
where the index $r$ specify inverted or repulsive, has attracted great
attention over the years in quantum mechanics \cite{Barton}-\cite{yuce2007}.
It is of course not positive definite but symmetric on the domain
$\mathcal{D\subset}$ in the Hilbert space $\mathcal{H}$ endowed with scalar
product $\left\langle .,.\right\rangle $ and related norm $\left\Vert
.\right\Vert $. It is well known that this system gives rise to a complex
eigenvalues and a non physical eigenvectors although $H^{\mathrm{r}}$ is
self-adjoint - the physical reason is that its potential is unbounded from below.

The inverted harmonic oscillator,obtained by changing $\omega$ to $\pm
i\omega$ in the harmonic oscillator, is often encountered as a
phenomenological model in dessipative processes and has been used among
others, as a model of instability in quantum mechanics \cite{baskoutas93}%
-\cite{shim2000}. In fact, its associated time-dependent solutions of \ the
Schr\"{o}dinger equation can be mapped to those of the Schr\"{o}dinger
equation for the harmonic potentials \cite{pedro2004}-\cite{Rajeev}. It yield
some physical informations on scales that span classical mechanics to quantum
field theory \cite{Mo}-\cite{Bha}. Recently in the Ref \ \cite{versha} , the
inverted harmonic oscillator Hamiltonian is used to understand the quantum
mechanics of scattering and time-decay in a diverse set of physical systems.

It have a wide range of application in different branches of physics
\cite{guth}-\cite{gai} such as, the tunneling effects , the mechanism of
matter-wave bright solitons, the cosmological model, and the\emph{ }quantum
theor\emph{y }of measurements.

The fact is that we are able to "invert" a standard oscillator by redefining
the frequency $\omega$.We will expect both of the normal and the inverted
oscillator representations to possess the same observable consequences.
However, it is not clear how this result truly comes about. It is important to
note that the regular and the inverted harmonic oscillators are genuinely
different. As a result, one cannot take for granted the known formulae
of\textrm{ }the regular oscillator and extrapolate them to the inverted
oscillators by simply replacing $\omega$ with $\pm i\omega$.\ Among the goals
of this paper is to define the transformation which connects the harmonic
oscillator to the inverted one and describes the physical situation, there by
clarifying several questions.

The analytic continuation of the angular velocity $\omega\rightarrow\pm
i\omega$ leads to an imaginary energy spectrum which eventually could be
obtained as an eigenvalue of the anti-$\mathcal{PT}$-symmetric non-Hermitian
Hamiltonian $(\pm iH^{\mathrm{os}})$. The question that quickly comes to mind
is can we relate $H^{\mathrm{r}}$ and $(\pm iH^{\mathrm{os}})$? or precisely,
which method should be used to obtain, from the anti-$\mathcal{PT}$
non-Hermitian oscillator $(\pm iH^{\mathrm{os}})$, the inverted oscillator
$H^{\mathrm{r}}$ without recourse to the substitution of\textrm{ }$\omega$ by
$\pm i\omega$? We will try to answer these\textrm{ }questions in the next section.

Another important branch in studying quantum systems is the so-called
non-Hermitian quantum mechanics. In non-Hermitian quantum mechanics, it was
found that the criteria for a quantum Hamiltonian to have a real spectrum is
that it possesses an unbroken $\mathcal{PT}$-symmetry ($\mathcal{P}$ is the
space-reflection operator or parity operator, and $\mathcal{T}$ is the
time-reversal operator) \cite{Bender98,Bender2002}. The concept of
$\mathcal{PT}$-symmetry has\textrm{ }found applications in several areas of
physics \cite{Ru}-\cite{Such} . Due\textsl{ }to\textsl{ }the fact that\textrm{
}the energy spectrum of $(\pm iH^{\mathrm{os}})$ being completely imaginary
that changes the physical structure of the system. We recall that a
$\mathcal{PT}$-symmetric system can be transformed to an anti-$\mathcal{PT}%
$-symmetric one by the transformation $H^{\mathrm{os}}\rightarrow((\pm
iH^{\mathrm{os}})$ \cite{Ge}-\cite{Peng} .

In analogy with the $\mathcal{PT}$-symmetric case, we call the
anti-$\mathcal{PT}$-symmetry of Hamiltonian $H$ unbroken if all of the
eigenfunctions of $H$ are eigenfunctions of\emph{ }the $\mathcal{PT}$
operator, i.e. when the energy spectrum of $H$ is entirely imaginary
\cite{Kheniche}. An alternative approach exploring the basic structure
responsible of a non- Hermitian Hamiltonian is the notion of the
pseudo-Hermiticity introduced in Ref. \cite{Scholz}. Mostafazadeh
\cite{Mostafa1} pointed out that the condition for a Hamiltonian $H$ to be
$\mathcal{PT}$-symmetric can be understood more generally as a special case of
pseudo-Hermiticity. An operator $H$ is said to be pseudo-Hermitian or
quasi-Hermitian if it satisfies the following relation
\begin{equation}
H^{\dagger}=\eta H\eta^{-1}, \label{1'}%
\end{equation}
where the Hermitian operator $\eta=\rho^{+}\rho$ and the Dyson operator $\rho$
are linear and invertible. The pseudo-Hermiticity\textrm{ }links as well the
pseudo-Hermitian Hamiltonian $H$ with an equivalent Hermitian Hamiltonian $h$
\begin{equation}
h=\rho H\rho^{-1}. \label{ph}%
\end{equation}

In what follows, we only consider the anti-$\mathcal{PT}$ -symmetric
Hamiltonian case $(iH^{\mathrm{os}})$. Similarly, the case $(-iH^{\mathrm{os}%
})$ would serve equally well.

In the present paper, we generate from anti-$\mathcal{PT}$-symmetric
Hamiltonian\ $(iH^{\mathrm{os}})$ an inverted harmonic oscillator-type
Hamiltonian $H^{\mathrm{r}}$ which is a Hermitian Hamiltonian and thus its
solution. The paper is organized as follows. In Section 2, we recall briefly
some properties of the standard harmonic and inverted oscillators. We then
introduce an appropriate quantum scaling operator $\eta$ which link the
anti-$\mathcal{PT}$-symmetric Hamiltonian oscillator\ $(iH^{\mathrm{os}})$ to
the inverted oscillator Hamiltonian $H^{\mathrm{r}}$. We obtain the set of
solutions of the inverted harmonic oscillator and also to define the full
orthonormalization relation of the eigenstates of $H^{\mathrm{r}}$. This
procedure allows us to construct the ladder operators in the inverted harmonic
oscillator, in Section 3, where we will address the problem of construction of
generalized coherent states associated with the inverted
oscillator\ $H^{\mathrm{r}}$. We obtain the mean values of the space and
momentum operators in the generalized coherent states and furthermore we
calculate the corresponding Heisenberg uncertainty. An outlook over the main
results is given in the conclusion.

\section{Generalized eigenfunction of the inverted harmonic oscillator}

Let us first recall some properties of the harmonic oscillator which are
useful to introduce the inverted oscillator. The Hamiltonian of the harmonic
oscillator is :
\begin{equation}
H^{\mathrm{os}}=\frac{1}{2m}p^{2}+\frac{1}{2}m\omega^{2}x^{2}, \label{ho}%
\end{equation}
and its the normalized eigenfunctions are given by means of Hermite
polynomial
\begin{equation}
\psi_{n}^{\mathrm{os}}(x)=\frac{1}{\sqrt{2^{n}n!}}\left(  \frac{\omega m}%
{\pi\hbar}\right)  ^{\frac{1}{4}}\exp\left[  \frac{-\omega m}{2\hbar}%
x^{2}\right]  H_{n}\left(  \sqrt{\frac{\omega m}{\hbar}}x\right)  ,
\label{ho1}%
\end{equation}
therefore the spectrum is discrete and the eigenvalues of (\ref{ho}) are
non-negative real numbers which correspond to the quantized energy levels
\begin{equation}
E_{n}^{\mathrm{r}}=\hbar\omega(n+\frac{1}{2}),\text{ \ \ \ \ \ \ }\omega>0,
\label{ho2}%
\end{equation}
and the orthonormal condition is well preserved
\begin{equation}
\left\langle \psi_{m}^{\mathrm{os}}(x)\right.  \left\vert \psi_{n}%
^{\mathrm{os}}(x)\right\rangle =\int\psi_{m}^{\ast\mathrm{os}}(x)\psi
_{n}^{\mathrm{os}}(x)dx=\delta_{mn}\text{ }, \label{ho3}%
\end{equation}
naturally, the linear combination $\sum_{n}c_{n}\psi_{n}^{\mathrm{os}}(x)$ is
a square integrable function for $\sum_{n}\left\vert c_{n}\right\vert
^{2}<\infty$.

The Hamiltonian (\ref{r1}) is formally obtainable from the harmonic oscillator
by the following change $\omega\rightarrow\pm i\omega$ and it corresponds to
the Hamiltonian of the harmonic oscillator with purely imaginary frequency.
Therefore, this replacement transforms the eigenfunctions of the harmonic
oscillator (\ref{ho1}) into generalized eigenvectors of the inverted harmonic
oscillator (\ref{r1})%
\begin{equation}
\psi_{n}^{\mathrm{r}}(x)=\frac{1}{\sqrt{2^{n}n!}}\left(  \frac{i\omega m}%
{\pi\hbar}\right)  ^{\frac{1}{4}}\exp\left[  \frac{-i\omega m}{2\hbar}%
x^{2}\right]  H_{n}\left(  \sqrt{\frac{i\omega m}{\hbar}}x\right)  ,
\label{r2}%
\end{equation}%
\begin{equation}
\widetilde{\psi}_{n}^{\mathrm{r}}(x)=\frac{1}{\sqrt{2^{n}n!}}\left(
\frac{-i\omega m}{\pi\hbar}\right)  ^{\frac{1}{4}}\exp\left[  \frac{i\omega
m}{2\hbar}x^{2}\right]  H_{n}\left(  \sqrt{\frac{-i\omega m}{\hbar}}x\right)
, \label{r2"}%
\end{equation}
which should be proved as eigenvectors of the inverted harmonic oscillator
(\ref{r1}) with the discrete purely imaginary spectrum%
\begin{equation}
E_{n}^{\mathrm{r}}=\pm iE_{n}^{\mathrm{os}}=\pm i\hbar\omega\left(  n+\frac
{1}{2}\right)  . \label{r2'}%
\end{equation}
\textsl{ }

From the mathematical point of view, the solution of the inverted Hamiltonian
in (\ref{r1}) is similar to that of the harmonic oscillator. But these
transformations, applied to the potential of the harmonic oscillator, turn it
into the potential $-\frac{1}{2}m\omega^{2}x^{2}$ and the same happens with
the eigenvalues: they are transformed from discrete real eigenvalues to
discrete imaginary ones. Once again we see that these generalized
eigenfunctions cannot represent physical states. When calculating the squared
norm of these functions
\begin{align}
\left\langle \psi_{n}^{\mathrm{r}}\right.  \left\vert \psi_{n}^{\mathrm{r}%
}\right\rangle  &  =\frac{1}{2^{n}n!}\left(  \frac{\omega m}{\pi\hbar}\right)
^{\frac{1}{2}}%
{\displaystyle\int}
H_{n}\left(  \sqrt{\frac{-i\omega m}{\hbar}}x\right)  H_{n}\left(  \sqrt
{\frac{i\omega m}{\hbar}}x\right)  dx\neq1,\label{r4}\\
\left\langle \widetilde{\psi}_{n}^{\mathrm{r}}\right.  \left\vert
\widetilde{\psi}_{n}^{\mathrm{r}}\right\rangle  &  =\frac{1}{2^{n}n!}\left(
\frac{\omega m}{\pi\hbar}\right)  ^{\frac{1}{2}}%
{\displaystyle\int}
H_{n}\left(  \sqrt{\frac{i\omega m}{\hbar}}x\right)  H_{n}\left(  \sqrt
{\frac{-i\omega m}{\hbar}}x\right)  dx\neq1, \label{r4'}%
\end{align}
we see\textsl{ }that they diverge in the limit $\left\vert x\right\vert
\rightarrow\infty$ as $x^{2n}$. Therefore any non-trivial linear combination
does not yield a square integrable function.\ Clearly, $\psi_{n}^{\mathrm{r}}$
($\widetilde{\psi}_{n}^{\mathrm{r}}$) are not elements of $L^{2}(R).$

Finally, we are left with the following question: what is the relation between
the eigenfunctions $\psi_{n}^{\mathrm{os}}(x)$ (\ref{ho1}) and those of the
inverted oscillator $\psi_{n}^{\mathrm{r}}(x)$ (\ref{r2}) (or (\ref{r2"}))?.In
order to establish a connection between them, we write $\psi_{n}^{\mathrm{r}%
}(x)$ (\ref{r2}) as
\begin{align}
\psi_{n}^{\mathrm{r}}(x)  &  =\exp\left[  -\frac{\pi}{8}(xp+px)\right]
\left[  \frac{1}{\sqrt{2^{n}n!}}\left(  \frac{\omega m}{\pi\hbar}\right)
^{\frac{1}{4}}\exp\left[  -\frac{\omega m}{2\hbar}x^{2}\right]  H_{n}\left(
\sqrt{\frac{\omega m}{\hbar}}x\right)  \right] \nonumber\\
&  =\exp\left[  -\frac{\pi}{8}(xp+px)\right]  \psi_{n}^{\mathrm{os}%
}(x)=e^{i\frac{\pi}{8}}\psi_{n}^{\mathrm{os}}(xe^{i\frac{\pi}{4}}), \label{r5}%
\end{align}
and $\widetilde{\psi}_{n}^{\mathrm{r}}(x)$ (\ref{r2"}) in the form%
\begin{align}
\widetilde{\psi}_{n}^{\mathrm{r}}(x)  &  =\exp\left[  \frac{\pi}%
{8}(xp+px)\right]  \left[  \frac{1}{\sqrt{2^{n}n!}}\left(  \frac{\omega m}%
{\pi\hbar}\right)  ^{\frac{1}{4}}\exp\left[  -\frac{\omega m}{2\hbar}%
x^{2}\right]  H_{n}\left(  \sqrt{\frac{\omega m}{\hbar}}x\right)  \right]
\nonumber\\
&  =\exp\left[  \frac{\pi}{8}(xp+px)\right]  \psi_{n}^{\mathrm{os}%
}(x)=e^{-i\frac{\pi}{8}}\psi_{n}^{\mathrm{os}}(xe^{-i\frac{\pi}{4}}),
\label{r5'}%
\end{align}
the operator $\exp\left[  \frac{\pi}{8}(xp+px)\right]  $ defines a complex
squeezed operator, because if $\left\vert \psi\right\rangle $ is the
state\textsl{ }vector of a system then $\exp\left[  \frac{\pi}{8}%
(xp+px)\right]  \left\vert \psi\right\rangle $ represents the same system
compressed in position space by the factor $e^{-i\frac{\pi}{4}}$ and expanded
in momentum space by the factor $e^{+i\frac{\pi}{4}}$.

It can be easily shown that under this transformation the coordinate and
momentum operators change according to%
\begin{align}
\exp\left[  \frac{\pi}{8}(xp+px)\right]  x\exp\left[  -\frac{\pi}%
{8}(xp+px)\right]   &  =xe^{-i\frac{\pi}{4}},\nonumber\\
\exp\left[  \frac{\pi}{8}(xp+px)\right]  p\exp\left[  -\frac{\pi}%
{8}(xp+px)\right]   &  =pe^{i\frac{\pi}{4}}, \label{r7}%
\end{align}
thus
\begin{equation}
\exp\left[  \frac{\pi}{8}(xp+px)\right]  H^{\mathrm{r}}\exp\left[  -\frac{\pi
}{8}(xp+px)\right]  =\left(  iH^{\mathrm{os}}\right)  . \label{r8}%
\end{equation}

We now return\textsl{ }to Eqs. (\ref{r2"}), (\ref{r5}), (\ref{r5'}) \ and
notice that
\begin{equation}
\widetilde{\psi}_{n}^{\mathrm{r}}(x)=\exp\left[  \frac{\pi}{8}(xp+px)\right]
\psi_{n}^{\mathrm{os}}(x)=\exp\left[  \frac{\pi}{4}(xp+px)\right]  \psi
_{n}^{\mathrm{r}}(x). \label{r8'}%
\end{equation}

We observe that these two families of generalized eigenvectors $\psi
_{n}^{\mathrm{r}}(x)$\ and $\widetilde{\psi}_{n}^{\mathrm{r}}(x)$ have the
remarkable properties:

i)\ Formulas (\ref{r5}) and (\ref{r5'}) imply that they are conjugated to each
other:
\begin{equation}
\widetilde{\psi}_{n}^{\ast\mathrm{r}}(x)=\psi_{n}^{\mathrm{r}}(x), \label{con}%
\end{equation}
which implies that $\psi_{n}^{\mathrm{r}}(x)$\ and $\widetilde{\psi}%
_{n}^{\mathrm{r}}(x)$ are related by the time-reversal operator $T\psi
_{n}^{\mathrm{r}}(x)$ $=\widetilde{\psi}_{n}^{\mathrm{r}}(x).$

ii) They are bi-orthonormal
\begin{align}
\left\langle \widetilde{\psi}_{m}^{\mathrm{r}}\right.  \left\vert \psi
_{n}^{\mathrm{r}}\right\rangle  &  =\left\langle \psi_{m}^{\mathrm{r}}\right.
\left\vert \widetilde{\psi}_{n}^{\mathrm{r}}\right\rangle \nonumber\\
&  =\left\langle \psi_{n}^{\mathrm{os}}\right\vert e^{\left[  -\frac{\pi}%
{8}(xp+px)\right]  }e^{\left[  \frac{\pi}{8}(xp+px)\right]  }\left\vert
\psi_{n}^{\mathrm{os}}\right\rangle =\delta_{mn}, \label{bio}%
\end{align}
the generalized orthonormal relations (\ref{bio}) involve the eigenvectors
$\left\vert \psi_{n}^{\mathrm{r}}\right\rangle $ as well as $\left\vert
\widetilde{\psi}_{n}^{\mathrm{r}}\right\rangle .$

iii)\ they are bi-complete
\begin{align}
\overset{\infty}{\underset{0}{\sum}}\left\vert \psi_{n}^{\mathrm{r}%
}(x)\right\rangle \left\langle \widetilde{\psi}_{m}^{\mathrm{r}}(x^{\prime
})\right\vert  &  =\overset{\infty}{\underset{0}{\sum}}\left\vert
\widetilde{\psi}_{m}^{\mathrm{r}}(x)\right\rangle \left\langle \psi
_{n}^{\mathrm{r}}(x^{\prime})\right\vert \nonumber\\
&  =e^{\left[  \frac{\pi}{8}(xp+px)\right]  }\underset{n}{\sum}\left\vert
\psi_{n}^{\mathrm{os}}\right\rangle \left\langle \psi_{n}^{\mathrm{os}%
}\right\vert e^{\left[  -\frac{\pi}{8}(xp+px)\right]  }=\delta(x-x^{\prime}).
\label{com'}%
\end{align}

The proof follows immediately from orthonormality and completeness of
the\emph{ }oscillator's eigenfunctions $\psi_{n}^{\mathrm{os}}$.Indeed,for the
harmonic oscillator, the completeness relation is $\underset{n}{\sum
}\left\vert \psi_{n}^{\mathrm{os}}(x)\right\rangle \left\langle \psi
_{n}^{\mathrm{os}}(x^{\prime})\right\vert =\delta(x-x^{\prime}),$ where the
kets $\left\vert \psi_{n}^{\mathrm{os}}\right\rangle =\exp\left[  \frac{\pi
}{8}(xp+px)\right]  \left\vert \psi_{n}^{\mathrm{r}}\right\rangle $ form a
complete set of bases, consequently the bi-completeness relation (\ref{com'})
is verified. Notice that, the generalized bi-orthonormal relation (\ref{bio})
can be rewritten as:
\begin{equation}
\left\langle \widetilde{\psi}_{m}^{\mathrm{r}}\right.  \left\vert \psi
_{n}^{\mathrm{r}}\right\rangle =\left\langle \psi_{n}^{\mathrm{r}}\right\vert
\eta\left\vert \psi_{n}^{\mathrm{r}}\right\rangle =\delta_{mn}, \label{r8"}%
\end{equation}
which introduces the notion of the $\eta$-pseudo-scalar product where the
operator $\eta$ is defined as%
\begin{equation}
\eta=\exp\left[  \frac{\pi}{4}(xp+px)\right]  =\rho^{+}\rho, \label{r9}%
\end{equation}
and
\begin{equation}
\rho=\exp\left[  \frac{\pi}{8}(xp+px)\right]  . \label{r10}%
\end{equation}

Now, the condition (\ref{r4}) yields%
\begin{equation}
\left\langle \psi_{m}^{\mathrm{r}}\right\vert \eta\left\vert \psi
_{n}^{\mathrm{r}}\right\rangle =%
{\displaystyle\int}
\psi_{n}^{\mathrm{os}}(x)\exp\left[  -\frac{\pi}{8}(xp+px)\right]  \text{
}\left[  \eta\right]  \text{ }\exp\left[  -\frac{\pi}{8}(xp+px)\right]
\psi^{\mathrm{os}}(x)dx=\delta_{mn}, \label{r11}%
\end{equation}
under this observation, we deduce that to insure the normalization condition
concerning the inverted eigenfunctions, one must clearly apply the $\eta$
pseudo-inner product $\left\langle ,\right\rangle _{\eta}$. This proves that
one of the fundamental\textsl{ }basis in the study of the inverted oscillator
is the pseudo-Hermicity concept.

Therefore, the operator $\eta$ links $(iH^{\mathrm{os}})$ to its hermitian
conjugate%
\begin{equation}
(-iH^{\mathrm{os}})=\eta(iH^{\mathrm{os}})\eta^{-1},
\end{equation}
which is nothing other than the quasi-Hermiticity relation. It then follows
immediately, that the two Hamiltonians $H^{\mathrm{r}}$ and $(iH^{\mathrm{os}%
})$ are related to each other as
\begin{equation}
H^{\mathrm{r}}=\rho^{-1}(iH^{\mathrm{os}})\rho.
\end{equation}

Therefore, Eq. (\ref{r11}) stands for the $\eta$ inner product $\left\langle
,\right\rangle _{\eta}$ in the pseudo-Hermitian case. The eigenfunctions of
the inverted harmonic oscillator $\psi_{n}^{\mathrm{r}}(x)$ are related to the
eigenfunctions of the harmonic oscillator $\psi_{n}^{\mathrm{os}}(x)$ via the
transformation operator $\rho$ (\ref{r10}) as
\begin{equation}
\psi_{n}^{\mathrm{r}}(x)=\rho^{-1}\psi^{\mathrm{os}}(x), \label{rhopsi}%
\end{equation}
thus any non-trivial linear combination $\underset{n}{\sum}c_{n}\psi
_{n}^{\mathrm{r}}(x)$ yields a square integrable function for $\underset
{n}{\sum}\left\vert c_{n}\right\vert ^{2}<\infty$.

As advertised before, let us back to the properties\textrm{ }(\ref{bio}) and
(\ref{com'})\textrm{ }of the generalized eigenvectors $\psi_{n}^{\mathrm{r}%
}(x)$\bigskip\ and $\widetilde{\psi}_{n}^{\mathrm{r}}(x)$\textrm{. }The
unbounded operator\textit{ }$\eta=\exp\left[  \frac{\pi}{4}(xp+px)\right]
$\textrm{ , }that cannot be defined on all of Hilbert space $\mathcal{H}%
$,\textrm{ }act on $\mathcal{H}\ $with domain $\mathfrak{D}$($\eta$) and
let\textit{ }$\mathcal{D}$\textit{ }be a dense subspace of\textit{
}$\mathcal{H}$ such that \textit{ }$\eta\mathcal{D}$\textit{ }$\subseteq
$\textit{ }$\mathcal{D}$, where $\mathcal{D}\subseteq\mathfrak{D}$($\eta$) $.$
We can define in $\mathcal{D}$\textit{ }the vectors $\left\vert \psi_{_{n}%
}\right\rangle $ and $\left\vert \phi_{_{n}}\right\rangle =\eta\left\vert
\psi_{_{n}}\right\rangle $ and the related sets $F_{\psi_{n}^{\mathrm{r}}}$ =
\{ $\left\vert \psi_{n}^{\mathrm{r}}\right\rangle $, $n\geq$ $0$\},
$F_{\widetilde{\psi}_{n}^{\mathrm{r}}}$ = \{ $\left\vert \widetilde{\psi}%
_{n}^{\mathrm{r}}\right\rangle =\eta\left\vert \psi_{n}^{\mathrm{r}%
}\right\rangle $, $n\geq$ $0$ \}. In particular, this lead to Eq. (\ref{r8"})
so that $F_{\psi_{n}^{\mathrm{r}}}$ and $F_{\widetilde{\psi}_{n}^{\mathrm{r}}%
}$ are biorthogonal sets and consequently $F_{\psi_{n}^{\mathrm{r}}}$ = \{
$\left\vert \psi_{n}^{\mathrm{r}}\right\rangle $\} and $F_{\widetilde{\psi
}_{n}^{\mathrm{r}}}$ = \{ $\left\vert \widetilde{\psi}_{n}^{\mathrm{r}%
}\right\rangle $ \} are bases for $\mathcal{H}.$

We can assert that the generalized orthonormal condition (\ref{bio}) or rather
(\ref{r8"}) can be interpreted as a biorthonormalization condition. Thus, from
the Eqs. (\ref{r5}) and (\ref{r8'}), the completeness relation (\ref{com'})
can be immediately deduced from that of the oscillator eigenfunction
,i.e,\textrm{ }$\underset{n}{\sum}\left\vert \psi_{n}^{\mathrm{os}%
}\right\rangle \left\langle \psi_{n}^{\mathrm{os}}\right\vert =\mathbf{I}$
\textrm{.}

Before abording, in the next section the coherent states of the inverted
oscillator $H^{\mathrm{r}}$, let us introduce the dimensionless
annihilation\emph{ }and\emph{ }creation operators of the quantum harmonic
oscillator $H^{\mathrm{os}}$as
\begin{equation}
a=\sqrt{\frac{m\omega}{2\hbar}}x+i\frac{p}{\sqrt{2m\hbar\omega}},\text{
\ \ }a^{+}=\sqrt{\frac{m\omega}{2\hbar}}x-i\frac{p}{\sqrt{2m\hbar\omega}},
\label{a}%
\end{equation}
these ladder operators (\ref{a}) can be represented in terms of
position\textrm{ }$x$\textrm{ }and momentum\textrm{ }$p$\textrm{
}operators\textrm{ }as%
\begin{equation}
x=\sqrt{\frac{\hbar}{2m\omega}}(a^{+}+a),\text{ \ \ \ \ }p=i\sqrt{\frac{\hbar
m\omega}{2}}(a^{+}-a), \label{x'}%
\end{equation}
while their commutation relation is%
\begin{equation}
\lbrack a,a^{+}]=1. \label{a'}%
\end{equation}

The best way to present the inverted coherent states is to translate their
definitions into the language of the coherent states of the harmonic
oscillator. Coherent states, or semi-classic states, are remarkable quantum
states that were originally introduced in 1926 by Schr\"{o}dinger for the
Harmonic oscillator \cite{Schrodinger} where the mean values of the position
and momentum operators in these states have properties close to the classical
values of the position $x_{c}(t)$ and the momentum $p_{c}(t).$ In particular,
we can construct coherent states of the harmonic oscillator $\left\vert
\alpha\right\rangle $ \cite{Glauber}-\cite{Sudarshan} as eigenstates of the
annihilation operator $a$%
\begin{equation}
a\left\vert \alpha\right\rangle =\alpha\left\vert \alpha\right\rangle ,\text{
\ \ }\alpha\in C. \label{2.1}%
\end{equation}

They\textsl{ }can be also obtained from the vacuum state $\left\vert
0\right\rangle $ by the action of the unitary displacement operator $D\left(
\alpha\right)  $
\begin{equation}
D\left(  \alpha\right)  \left\vert 0\right\rangle =\exp\left(  \alpha
\text{\ }a^{+}-\alpha^{\ast}a\right)  \left\vert 0\right\rangle . \label{2.4}%
\end{equation}

\section{Generalized coherent states for the inverted harmonic oscillator}

One can verify that in the case of\emph{ }the\emph{ }inverted oscillator, the
form of Hamiltonian (\ref{r1}) reads
\begin{equation}
H^{\mathrm{r}}=\frac{i\hbar\omega}{2}(\mathcal{\bar{A}}\text{ }\mathcal{A}%
+\mathcal{A}\text{ }\mathcal{\bar{A}}), \label{Hr3}%
\end{equation}
where the ladder operators $\left(  \mathcal{A},\mathcal{\bar{A}}\right)  $
are linked to those in Eq.(\ref{a}) through the transformation
\begin{equation}
\mathcal{A}=\rho^{-1}a\rho=\exp\left[  i\frac{\pi}{4}\right]  \left(
\sqrt{\frac{m\omega}{2\hbar}}x+\frac{p}{\sqrt{2m\omega\hbar}}\right)  =\left(
\sqrt{\frac{im\omega}{2\hbar}}x+\frac{ip}{\sqrt{2mi\omega\hbar}}\right)  ,
\label{A1}%
\end{equation}%
\begin{equation}
\mathcal{\bar{A}}=\rho^{-1}a^{\dagger}\rho=\exp\left[  i\frac{\pi}{4}\right]
\left(  \sqrt{\frac{m\omega}{2\hbar}}x-\frac{p}{\sqrt{2m\omega\hbar}}\right)
=\left(  \sqrt{\frac{mi\omega}{2\hbar}}x-\frac{ip}{\sqrt{2mi\omega\hbar}%
}\right)  , \label{A2}%
\end{equation}
and satisfy the following commutation relation $\left[  \mathcal{A}%
,\mathcal{\bar{A}}\right]  $ $=1.$ As the case of the harmonic oscillator, the
coherent states for the inverted oscillator $\left\vert \varphi_{\alpha
}^{\mathrm{r}}\right\rangle ^{\mathcal{A}}$ can be defined as eigenstates of
the annihilation operator $A$\
\begin{equation}
\mathcal{A}\left\vert \varphi_{\alpha}^{\mathrm{r}}\right\rangle
^{\mathcal{A}}=\alpha\left\vert \varphi_{\alpha}^{\mathrm{r}}\right\rangle
^{\mathcal{A}},\text{ \ \ \ \ \ }\alpha\in C. \label{cs}%
\end{equation}

It should be noted that Eqs. (\ref{A1}) and (\ref{A2}) indicate that the
eigenvalue $\alpha$ can be considered as being a real eigenvalue of the
Hermitian operator $\left(  \sqrt{\frac{m\omega}{2\hbar}}x+\frac{p}%
{\sqrt{2m\omega\hbar}}\right)  $ multiplied by the phase factor $e^{+i\frac
{\pi}{4}}:$
\begin{equation}
\alpha=\left\vert \alpha\right\vert e^{+i\frac{\pi}{4}}. \label{alpha}%
\end{equation}

In the $x$ representation, Eq. (\ref{cs}) can be written as
\begin{equation}
\left(  \sqrt{\frac{mi\omega}{2\hbar}}x+\frac{\hbar}{\sqrt{2mi\omega\hbar}%
}\frac{\partial}{\partial x}\right)  \varphi_{\alpha}^{\mathrm{r}}%
(x)=\alpha\varphi_{\alpha}^{\mathrm{r}}(x), \label{c1}%
\end{equation}
and can be explicitly solved; we obtain
\begin{align}
\varphi_{\alpha}^{\mathrm{r}}(x)  &  =\left(  \frac{im\omega}{2\hbar\pi^{2}%
}\right)  ^{\frac{1}{4}}e^{-\frac{i}{2}\left\vert \alpha\right\vert ^{2}}%
\exp\left[  \left(  \sqrt{\frac{2im\omega}{\hbar}}\alpha x-i\frac{m\omega
}{2\hbar}x^{2}\right)  \right] \nonumber\\
&  =\left(  \frac{im\omega}{2\hbar\pi^{2}}\right)  ^{\frac{1}{4}}e^{-\frac
{i}{2}\left\vert \alpha\right\vert ^{2}}\exp\left[  \alpha\left(
\mathcal{A}+\mathcal{\bar{A}}\right)  \right]  \exp\left[  -i\frac{m\omega
}{2\hbar}x^{2}\right] \nonumber\\
&  =\left(  \frac{im\omega}{2\hbar\pi^{2}}\right)  ^{\frac{1}{4}}\exp\left[
\alpha\mathcal{\bar{A}}\right]  \exp\left[  \alpha\mathcal{A}\right]
\exp\left[  -i\frac{m\omega}{2\hbar}x^{2}\right] \nonumber\\
&  =\exp\left[  \alpha\mathcal{\bar{A}}\right]  \varphi_{0}^{\mathrm{r}}(x),
\label{c2'}%
\end{align}
where the vacuum state of the\ inverted oscillator $\varphi_{0}^{\mathrm{r}%
}(x)=\left(  \frac{im\omega}{2\hbar\pi^{2}}\right)  ^{\frac{1}{4}}\exp\left[
-i\frac{m\omega}{2\hbar}x^{2}\right]  $ is not square integrable. The vacuum
states $\left\{  \left\vert \varphi_{0}^{\mathrm{r}}\right\rangle ,\left\vert
\widetilde{\varphi}_{0}^{\mathrm{r}}\right\rangle \right\}  $ $\in$ $D$ are
related to each other as $\left\vert \widetilde{\varphi}_{0}^{\mathrm{r}%
}\right\rangle =\eta\left\vert \varphi_{0}^{\mathrm{r}}\right\rangle $. Thus,
the biorthonormalization condition $\left\langle \widetilde{\varphi}%
_{0}^{\mathrm{r}}(x)\right.  \left\vert \varphi_{0}^{\mathrm{r}}%
(x)\right\rangle =\left\langle \varphi_{0}^{\mathrm{r}}(x)\right\vert
\eta\left\vert \varphi_{0}^{\mathrm{r}}(x)\right\rangle =I$ is verified.

When the inverted oscillator is in a particular state $\varphi_{\alpha
}^{\mathrm{r}}(x)$ (\ref{c2'}) at the instant $t=0$,\textrm{ }how do its
physical properties evolve over time?

\ Suppose that a system is in the state $\varphi_{\alpha}^{\mathrm{r}}(x)$ at
$t=0$, and when the Hamiltonian is not time-dependent then its state at all
$t$ will be given by :%
\begin{align*}
\psi_{\alpha}^{\mathrm{r}}(x,t)  &  =\exp\left[  -\frac{i}{\hbar}%
H^{\mathrm{r}}t\right]  \varphi_{\alpha}^{\mathrm{r}}(x)\\
&  =\exp\left[  -iH^{\mathrm{r}}t\right]  \exp\left[  \alpha\mathcal{\bar{A}%
}\right]  \varphi_{0}^{\mathrm{r}}(x),
\end{align*}
therefore we write
\begin{equation}
\exp\left[  -\frac{i}{\hbar}H^{\mathrm{r}}t\right]  \varphi_{\alpha
}^{\mathrm{r}}(x)=\exp\left[  -\frac{i}{\hbar}H^{\mathrm{r}}t\right]
\exp\left[  \alpha\mathcal{\bar{A}}\right]  \exp\left[  \frac{i}{\hbar
}H^{\mathrm{r}}t+\frac{\omega}{2}t\right]  \varphi_{0}^{\mathrm{r}}(x),
\end{equation}
where we used $\left(  H^{\mathrm{r}}-i\frac{\omega\hbar}{2}\right)
\varphi_{0}^{\mathrm{r}}(x)=0$.

Knowing that
\[
\left[  -\frac{i}{\hbar}H^{\mathrm{r}}t,\alpha\mathcal{\bar{A}}\right]
=\omega t\alpha\mathcal{\bar{A}},
\]
we can thus use a special case of the Baker-Campbell-Hausdorff formula which
states that if $\ [a,B]=cB$ with $c\in C$, then
\begin{equation}
\exp[{\LARGE a}]\exp[B]\exp[-{\LARGE a}]=\exp[\exp(c)B]. \label{4}%
\end{equation}

We apply eq. (\ref{4}) with $a=-\frac{i}{\hbar}H^{\mathrm{r}}t$,
$B=\alpha\left(  \mathcal{\bar{A}}\right)  $ and $c=\omega t$ to obtain%
\begin{equation}
\psi_{\alpha(t)}^{\mathrm{r}}(x,t)=\exp\left[  \frac{\omega}{2}t\right]
\exp\left[  e^{\omega t}\alpha\mathcal{\bar{A}}\right]  \varphi_{0}%
^{\mathrm{r}}(x)=\exp\left[  \frac{\omega}{2}t\right]  \varphi_{\alpha
}^{\mathrm{r}}(x), \label{ev}%
\end{equation}
if we compare this result with (\ref{c2'}), we see that, to go from
$\psi_{\alpha}^{\mathrm{r}}(x,0)=$ $\varphi_{\alpha}^{\mathrm{r}}(x)$ to
$\psi_{\alpha}^{\mathrm{r}}(x,t)$,\ all we must do is to change $\alpha$ to
$\alpha(t)=\alpha e^{\omega t}$ and multiply the obtained state by
$\exp\left[  \frac{\omega}{2}t\right]  $\ (which is a global amplitude factor
). We already know, for the harmonic oscillator, that the mean values
$\left\langle x\right\rangle ^{\mathrm{os}}(t)=x_{c}^{\mathrm{os}}(t)$ and
$\left\langle p\right\rangle ^{\mathrm{os}}(t)=-m\omega x_{c}^{\mathrm{os}%
}(t)$ always remain equal to the corresponding classical values. What about
the mean values of physical quantities of the inverted oscillator?

The mean values $\left\langle x\right\rangle _{\eta}^{\mathrm{r}}$ and
$\left\langle p\right\rangle _{\eta}^{\mathrm{r}}$ can be obtained by
expressing $x$ and $p$ in terms of $\mathcal{A}$ and $\mathcal{\bar{A}}$
[Eqs.( \ref{A1}) and (\ref{A2})]%
\begin{equation}
x=\sqrt{\frac{\hbar}{2mi\omega}}(\mathcal{A}+\mathcal{\bar{A}}),\text{
\ \ \ \ }p=i\sqrt{\frac{\hbar mi\omega}{2}}(\mathcal{\bar{A}-A}),
\end{equation}
we see that
\begin{equation}
\left\langle x\right\rangle _{\eta}^{\mathrm{r}}=\sqrt{\frac{\hbar}{2mi\omega
}}\left\langle \psi_{\alpha}^{\mathrm{r}}(x,t)\right\vert \eta(\mathcal{A}%
+\mathcal{\bar{A}})\left\vert \psi_{\alpha}^{\mathrm{r}}(x,t)\right\rangle
=\sqrt{\frac{2\hbar}{m\omega}}\left\vert \alpha\right\vert e^{\omega
t}=x_{\eta}^{\mathrm{r}}(t),
\end{equation}%
\begin{equation}
\left\langle p\right\rangle _{\eta}^{\mathrm{r}}=i\sqrt{\frac{\hbar mi\omega
}{2}}\left\langle \psi_{\alpha}^{\mathrm{r}}(x,t)\right\vert \eta
(\mathcal{\bar{A}-A})\left\vert \psi_{\alpha}^{\mathrm{r}}(x,t)\right\rangle
=\sqrt{2\hbar m\omega}\left\vert \alpha\right\vert e^{\omega t}=p_{\eta
}^{\mathrm{r}}(t),
\end{equation}
the time dependence of the expectation value matches the classical one of the
inverted oscillator.

An analogous calculation yields:
\begin{equation}
\left\langle x^{2}\right\rangle _{\eta}^{\mathrm{r}}=\frac{\hbar}{2m\omega
}\left[  \left(  \alpha(t)+\alpha^{\ast}(t)\right)  ^{2}+1\right]  \text{,
\ \ \ \ \ }\left\langle p^{2}\right\rangle _{\eta}^{\mathrm{r}}=\frac{\hbar
m\omega}{2}\left[  1-\left(  \alpha(t)-\alpha^{\ast}(t)\right)  ^{2}\right]  ,
\end{equation}
and therefore:%
\begin{equation}
\left(  \Delta x\right)  _{\eta}^{\mathrm{r}}=\sqrt{\frac{\hbar}{2m\omega}%
},\ \ \ \ \ \left(  \Delta p\right)  _{\eta}^{\mathrm{r}}=\sqrt{\frac{\hbar
m\omega}{2}}.
\end{equation}

Neither $\left(  \Delta x\right)  _{\eta}^{\mathrm{r}}$ nor $\left(  \Delta
p\right)  _{\eta}^{\mathrm{r}}$ depends on $\alpha$. Note also that $\left(
\Delta x\right)  _{\eta}^{\mathrm{r}}.\left(  \Delta p\right)  _{\eta
}^{\mathrm{r}}$ takes on its minimum value:
\[
\left(  \Delta x\right)  _{\eta}^{\mathrm{r}}.\left(  \Delta p\right)  _{\eta
}^{\mathrm{r}}=\frac{\hbar}{2}%
\]

\section{Conclusion}

It is well known that the system described by the inverted oscillator gives
rise to the generalized complex eigenvalues. The physical reason for that is
the potential being unbounded from below. The corresponding energy eigenstates
for (\ref{r1}) have been found in terms of the parabolic cylinder functions
$D_{\nu}(x)$ \cite{chru2004}.

Using a nonconventional technique, i.e. a quasi-Hermiticity in quantum
mechanics, we have solved the problem relative to the inverted oscillator. In
fact, we find that the system can be entirely expressed in terms of idealized
states that are related to those of the harmonic oscillator. The connection
with a harmonic oscillator may be established by the scaling operator $\rho$
(\ref{r10}).

Ideal physical systems are conceptually Hermitian, but realistic systems are
sometimes non-Hermitian because of their interactions with their environments.
We have given the principal properties of the harmonic oscillator
$H^{\mathrm{os}}$ formulation in quantum mechanics and we introduced the
inverted oscillator $H^{\mathrm{r}}$. Using the notion of anti-$\mathcal{PT}$
-symmetric non-Hermitian Hamiltonian, this led us to show that $H^{\mathrm{r}%
}$, $\left(  iH^{\mathrm{os}}\right)  $ and their two sets of eigenfunctions
are connected in a simple manner. Therefore, the eigenfunctions of the
inverted oscillator are pseudo-orthonormal.

The coherent states of the ordinary oscillator are special wave groups giving
probability distributions whose shapes never change, and follow classical
trajectories. They are obtained as eigenstates of the annihilation operator,
or equivalently by acting on vacuum eigenstate with the unitary displacement
operator $D\left(  \alpha\right)  $ introduced in section 2. One might think
that the coherent states for the inverted oscillator seem less important,
largely because, unlike groups of waves, they are not fully integrable which
is not at all appropriate.

So, to show their importance, we addressed the problem of construction of
ladder operators for $H^{\mathrm{r}}$ and their associated integrable coherent
states. To do this, we took as reference state the vacuum state of \ the
inverted oscillator and, as in the case of the harmonic oscillator, the
generalized coherent states for inverted oscillator are defined as eigenstates
of the annihilation operator $A=\rho^{-1}a\rho$ where $a$ is the annihilation
operator associated to the harmonic oscillator. These generalized coherent
states can be also obtained by the action on the vacuum state of the inverted
oscillator of a non unitary displacement operator.

Then, we showed that the mean values of the position and momentum operators in
these coherent states have properties close to the classical values of the
position $x_{c}^{\mathrm{r}}(t)$ and the momentum $p_{c}^{\mathrm{r}}(t)$. The
corresponding Heisenberg\ uncertainty relation is minimum.

\paragraph{ACKNOWLEDGMENTS}

The authors are grateful to Prof A. Layadi (UFAS)\ and Dr N. Mana for the
valuable suggestions that led to considerable improvement in the presentation
of this paper.

\end{document}